# Comparative Analysis of Network Forensic Tools and Network Forensics Processes

Fahad M Ghabban
College of Computer Science and Engineering. Information system department. Taibah University. Saudi Arabia. Madina
fghaban@taibahu.edu.sa

Ibrahim Alfadli
College of Computer Science and Engineering. Information system department. Taibah University. Saudi Arabia. Madina
ialfadli@taibahu.edu.sa

Omair Ameerbakhsh
College of Computer Science and Engineering. Information system department. Taibah University. Saudi Arabia. Madina
oameerbakhsh@taibahu.edu.sa

Amer Nizar AbuAli
College of Computer Science and Engineering. Information system department. Taibah University. Saudi Arabia. Madina
aabuali@taibahu.edu.sa

Arafat Al-Dhaqm
Faculty of Engineering, School of Computing,
University Technology Malaysia
Malaysia, Johor
mrarafat1@utm.my

Mahmoud Ahmad Al-Khasawneh
Faculty of Computer & Information Technology
Al-Madinah International University
Shah Alam, Malaysia
mahmoud@outlook.my

*Abstract*— Network Forensics (NFs) is a branch of digital forensics which used to detect and capture potential digital crimes over computer networked environments crime. Network Forensic Tools (NFTs) and Network Forensic Processes (NFPs) have abilities to examine networks, collect all normal and abnormal traffic/data, help in network incident analysis, and assist in creating an appropriate incident detection and reaction and also create a forensic hypothesis that can be used in a court of law. Also, it assists in examining the internal incidents and exploitation of assets, attack goals, executes threat evaluation, also by evaluating network performance. According to existing literature, there exist quite a number of NFTs and NTPs that are used for identification, collection, reconstruction, and analysing the chain of incidents that happen on networks. However, they were vary and differ in their roles and functionalities. The main objective of this paper, therefore, is to assess and see the distinction that exist between Network Forensic Tools (NFTs) and Network Forensic Processes (NFPs). Precisely, this paper focuses on comparing among four famous NFTs: Xplico, OmniPeek, NetDetector, and NetIetercept. The outputs of this paper show that the Xplico tool has abilities to identify, collect, reconstruct, and analyse the chain of incidents that happen on networks than other NF tools.

## I. Introduction

Digital forensics as a forensic discipline has evolved to encompass diverse subdomains such as computer forensics, mobile device and small device forensics, software forensics, multimedia forensics, as well as network forensics[1]–[8] [9] Network Forensic Tools (NFTs) are able to collect the whole network stream of traffic, permit customers to analyse the network stream of traffic based on their requirements and find the main elements about the traffic [10], [11]. NFTs are able to be aligned with IDSs and firewalls in order to create the protection of network, also to record all traffic records for rapid examination. NFTs allow corporations of seized, collected, and analysed packets of network traffic, which allows the investigator to obtain traffic forms between different machines. Several NFTs have been offered in the literature which offer consistent data collection and strong analytical abilities. Furthermore, numerous security tools are proposed for network security. However, these tools have been created for collecting and processing evidence, and do not have a forensic purpose [12], [13]. This paper focuses on comparing among five NFTs: Xplico, OmniPeek, NetDetector, and NetIetercept. Then select the best tool based on their capabilities and features.

The structure of this paper arranged as follows: Section 2 presents the digital forensic field. Section 3 reviews the network forensic tools and Section 4 offers the discussion part, and Section 5 concludes the paper.

## II. DIGITAL FORENSICS

The digital forensics discipline is used to investigate cybercrimes [7], [14], [15][16]. It consists of several branches as shown in Figure 1: Database forensic, mobile forensics, computer forensics, social network forensics, forensic data analysis, IoT forensics [17], drones forensic, and network forensics. The database forensics field is a branch of digital forensics which used to solve database crimes. It received much attention from researchers to address database incidents [18]–[26]. The database forensic field is complex and heterogeneous field due to the variety of database system infrastructures and m multidimensional nature of the database systems [27]–[29]. Therefore, at the time of writing this article, this domain lacks a unified forensic framework to facilitate and organize domain knowledge amongst domain forensic users [30]–[32]. Additionally, the database forensic domain lacks a universal forensic tool which creates many challenges among domain practitioners [33], [34]. Mobile forensics is a significant branch of digital forensics which used to address mobile cybercrimes. It received many works to overcome mobile issues [8], [35]–[37]. However, it is suffering from several drawbacks, such as redundant investigation terminologies, processes, concepts, and practices. On the contrary, computer forensics is a branch of Digital Forensics (DF) that used to detect, collect, preserve, and analyse evidence data [38]–[40]. NF on the other hand is a branch of digital forensics which aims to reveal network cybercrimes. Thus, this paper focuses on NFTs.

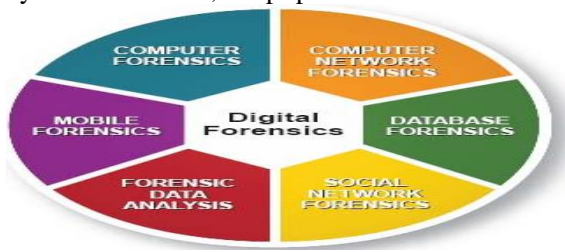

*Fig. 1. Digital forensics fields*

### III. NETWORK FORENSICS FIELD

Several investigation processes models have been proposed for NF in the literature. However, these models are proposed for specific purposes. For example, [41] proposed a new common process model and analyses several operations for NF as shown in Figure 2. They produced a new framework to solve the research limitations and review the work-in-progress. The proposed framework consists of four investigation processes: multi-sensor data fusion, identification of attack events, attack reconstruction, and incident response. Multi-sensor data fusion confirms that the complete incident/attack data is seized for investigation of the incident/attack. Identification of helpful network actions facilitates data decrease as unnecessary data is deleted. Attack reconstruction and traceback allow the trial of the attacker. Incident response enables improvement and reduces destruction [42], [43][44].

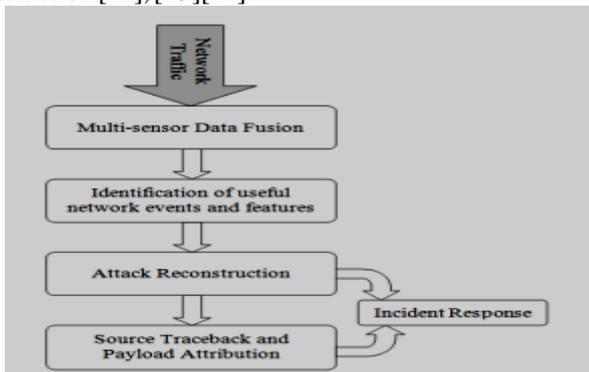

*Fig. 2. Common process model for NFs field* [41]

A common process framework for NF has been proposed by [45] as shown in Figure 3. It consists of nine investigation processes: preparation, detection, incident response, collection, preservation, examination, analysis, investigation, and presentation. The preparation process is used to prepare a mandatory authorization and screech warrant to avoid any illegal issues. The detection process is used to verify, check, and confirm the suspected attacks. The purpose of the incident response process is to put an action strategy of how to respond to future attacks and recover from the current destruction. The collection process is an important/significant process used to collect data from sensors, where is the preservation process is used to protect the integrity of the collected data [46][47]. The examination process is used to check and evaluate the authentication or originality of the collected data. The analysis process is used to reconstruct the timeline of the events and reveal the evidence. The investigation process is used to verify the path from a victim network or system via any intermediary systems and communication pathways, back to the point of attack origination. Leveraging human psychosocial attribute, studies in [42], [43], [48], [49] developed human centric network forensic frameworks. Their frameworks and model contain diverse human behavioural component which is capable of distinguishing human at the network level. Extending the human component, studies in [50]–[52] integrated behavioural component as a process for network attribution. Attempt to further finetune the attribution was carried out using digital forensic readiness process, ranging from the cloud [38], [53]–[56] to the intranet traffic [57]. The final process presentation is used to present whole investigation tasks in a human language for legal personal and providing a declaration of several processes used to arrive at the conclusion [58].

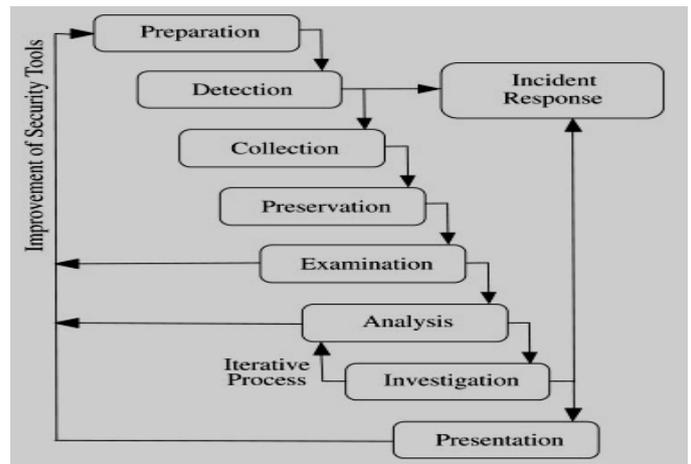

*Fig. 3. Generic process model for NFs field* [45]

A network forensic model for SYN attack has been proposed by [59] to offer effective preservation as shown in Figure 4. It consists of three investigation processes: collection and preservation module, analysis module, and preservation module. The collection and preservation process are used to seize whole network events from the network interface for forensic analysis. The analysis module is used to analyse the whole collected data started in the host system[13][60]. The purpose of the presentation module is to reveal the analysis module production if allowing port scanning attack [61], [62].

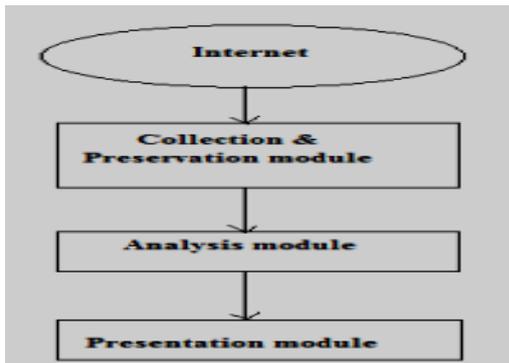

*Fig. 4. Architecture model for NFs field* [59]

A common model for could network forensics has been proposed by [63][15]. It consists of five investigation processes as illustrated in Figure 5: reporting, analysis, aggregation, separation, and data collection.

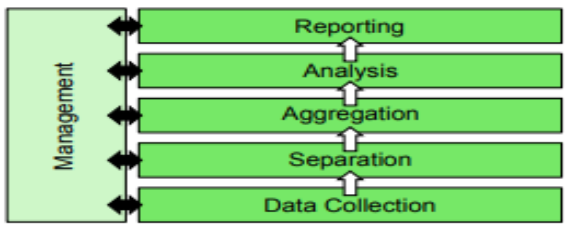

*Fig. 5. Cloud forensics process model* [64]

A common process for botnet forensics is presented by [64]. The specific limitation of the research which is existing in implementation is recognized and offered as challenges [56][65]. This model consists of nine investigation processes as displays in Figure 6: Preparation of security tools, Reorganizations of bots, incident response, collection, retention, inspection, analysis, investigation, and results.

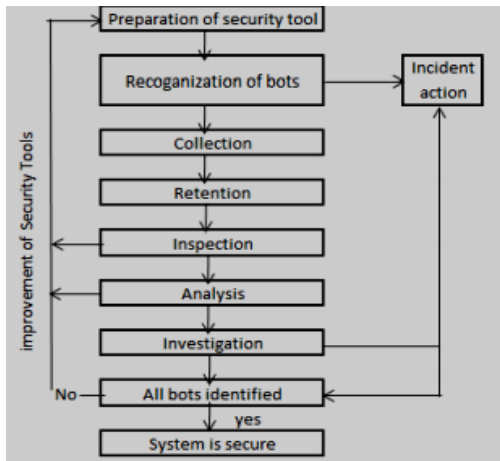

*Fig. 6. Botnet process model* [64]

A collection process model has been introduced by [66]. It consists of one collection process. The main purpose of this model is to collect evidence that can categorize the highly precise potential evidence [52]. Several process models have been proposed for the NFs field, however, it suffers from redundant investigation processes. For, example [41] consists of four (4) processes, [45] entails nine(9) investigation processes, [59] comprises of three investigation processes, [63] includes five (5) investigation processes, [64] involves nine (9) investigation processes, and finally [66] has one investigation process. Therefore, the NFs field needs a high abstract model (metamodel) to combine and unify whole existing process models [67].

IV. NETWORK FORENSIC TOOLS

NFTs have been proposed for NF in the literature as shown in Figure 2. However, these tools are proposed for specific purposes.

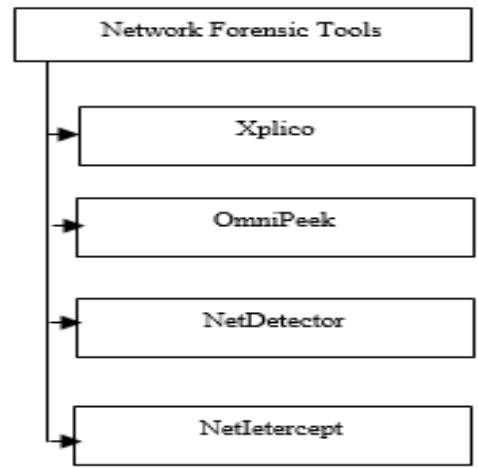

*Fig. 7. Network forensics tools*

For example, Xplico is an open-source network forensic analysis tool that is used to reconstructing audio, video, images, pdf, and several other text files from a network capture [65], [68], [69] It developed for the Linux platform. It has several capabilities:

1) *Collect application data.\*
2) *Collect information from database/files using SQLite or MYSQL*
3) *Analysis capabilities: Support online and offline analysis of packet capture.*
4) *Analyze live stream of traffic*
5) *Support many protocols ARP, PPP, VLAN, IPV4, IPV6, SNOOP, TCP, IRC, HYYP, SMTP, FTP, SIP, HTTP, DNS, and UDP.*
6) *Presentation and Reporting: Presented results in a human-readable manner and producing data in colorful*

   *tables.*
7) *Platform: Linux*
8) *Tool supporting: it supports most of the digital forensic fields*

The second NF tool is OmniPeek which is used to analyse collected packets [70], [71].[72] It was developed by WildPackets Inc. This tool has two major roles of network troubleshooting and procedure examination. It has many capabilities:

1) *Support email analysis sflow and netflow statistics*
2) *Platform: Linux and Windows.*
3) *GUI: easy to use ( fast examining, drill-down, and correcting performing jams through a variety of network)*
4) *Collect data from any network topology.*
5) *Support many borders and links to an unrestricted amount of TimeLine.*

The third NF tool is NetDetector which is one of the most famous NF tools [73]. This tool has been developed for security purposes [74]. It is used for the discovery of malicious activities in the network. It has many features:

1) *IDS tool*
2) *Analysing how the attack did happen and by whom, when the attack did happen, what the proper actions can be carried to fix the attack.*
3) *Analysing security protocols of the network.*

The fourth NF tool is NetIetercept which is used to collect and analyse network traffics for real-time catching [73]. It has many features:

1) *It works with UDP and TCP protocols.*
2) *Advanced examination skills*
3) *GUI that permits investigators to select proper features, and create complete reports, invisible to network users.*

Throughout this survey, it is very clear that the NF field has several forensic tools which differ in their roles and capabilities [75]. For example, the Xplico tool has many capabilities to collect. Analyse, and present network activities, however, it is specific for Linux OS, whereas OmniPeek tool is supported by two platforms (Linux and Windows), however, it is not a pure forensics tool [50]. It is used for network troubleshooting and procedure examination. Although, NetDetector is the most NF tool, however, it is developed for security purposes. It does not have the capabilities to analyse and present evidence. Therefore, the best NF tool amongst these four tools is Xplico which has several capabilities to identify, collect, examine, analyse, and present network crimes [76].

## V. DISCUSSION

The process of understanding the effectiveness and reliability of a network forensic tools is often hinged on the underlying model on which specific tool was developed. Whilst NetIetercept and NetDetector could provide a veritable tool for potential evidence acquisition: a phase of the network forensic models, they are largely limited to that phase of forensics. This, therefore, limits the potential effectiveness of the tools. Therefore, a phase specific tool would always be limited to the actual phase it is designed to address. Xplico, on the other hand, has the potential to span multiple phases of the network forensic model. Whilst the use is not limited to network forensic analysis, it has the potential to evolve towards a one-stop-shop to network forensics. This is similar to tools such as Autopsy, which offers open-source forensic capability. Furthermore, the logic of cross platform gives an added functionality for all forensic investigators. Going forward, therefore, network forensic researchers and practitioners can leverage the functionalities provided by this open-source tools for effective forensic investigation.

## VI. CONCLUSION

Several NFTs have been offered for the network forensic field to identify, capture, collect, analyse, and document network incidents. These tools are varying in their roles and functionalities. Thus, this paper, conducted a comparative analysis amongst four famous NFTs to select the best forensic tool. The outcome of this paper shows that the Xplico tool is the best which covers whole forensic tasks (identification, collection, examination, analysis, and presentation). The future work of this paper is to review more NFTs and conduct real scenario/case studies to evaluate the capabilities of the Xplico tool.